\def\etal{{\rm et al. }}
\def\mpc{{\ \rm Mpc}}
\def\kpc{{\ \rm kpc}}
\def\kms{{\ \rm km\, s^{-1}}}

\documentclass[usenatbib]{mn2e}

\usepackage{amssymb}
\usepackage{graphicx}
\usepackage{bmpsize}
\usepackage{cite}
\usepackage{natbib}
\usepackage{hyperref} 
\usepackage{aas_macros}
\usepackage{times}
\usepackage{amsmath}
\usepackage{natbib}
\usepackage{longtable}
\usepackage{txfonts}
\usepackage{pdflscape}
\usepackage{subfigure}
\usepackage{xcolor} 

\begin{document}

\title{The Global Environment of Small Galaxy Systems}
\author[Duplancic \etal]{Fernanda Duplancic$^{1}$\thanks{E-mail: fduplancic@unsj-cuim.edu.ar}, Federico D\'avila-Kurb\'an$^{2}$, Georgina V. Coldwell$^{1}$, Sol Alonso$^{1}$,
\newauthor
Daniela Galdeano$^{1}$\\ 
$^{1}$ Departamento de Geof\'{i}sica y Astronom\'{i}a, CONICET, Facultad de Ciencias Exactas, F\'{i}sicas y Naturales, Universidad Nacional \\
de San Juan, Av. Ignacio de la Roza 590 (O), J5402DCS, Rivadavia, San Juan, Argentina\\
$^{2}$ Instituto de Astronom\'ia Te\'orica y Experimental (IATE-CONICET), Laprida 854, C\'ordoba, Argentina }

\date{\today}

\pagerange{\pageref{firstpage}--\pageref{lastpage}}

\maketitle

\label{firstpage}

\begin{abstract}

The main goal of this work is to investigate the influence of environment at different scales on the properties of galaxies in systems with a low number of members. To this end we used a catalogue of small galaxy systems comprising compact and locally isolated pairs, triplets and groups with four and up to six galaxies. We consider fixed aperture estimators and found that at scales lower than 5$\mpc$ pairs are associated to lower density environments than triplets and groups. Moreover a nearest neighbour approach highlights that triplets prefer denser environments than pairs and slightly less dense environments than groups. When considering the position within the cosmic web we found that pairs and triplets in our sample are associated to void environments while galaxy groups are more likely to reside in void walls. In agreement with these results, the system-galaxy cross-correlation function shows that pairs inhabit environments of lesser density compared to triplets and groups, and on small scales ($< 3\mpc$) triplets appear to behave as an intermediate system. Related to the properties of neighbour galaxies of small systems we found that the neighbours of groups present a lower fractions of star forming, young stellar population and blue colour galaxies with respect to neighbours of triplet and pair systems. These results suggest that differences in the properties of galaxies in pairs, triplets and groups are not only related to the existence of an extra galaxy member but also to the large scale environment inhabited by the systems.
 
\end{abstract}

\begin{keywords}
galaxies: groups: general;
galaxies: interactions;
galaxies: statistics
\end{keywords}

\section{Introduction}
\label{intro}
In the standard scenario of hierarchical clustering in the Universe, the study of the environment is crucial to understand galaxy evolution. It has been shown that several properties of galaxies have a strong correlation with environment \citep{Dressler1980,Kauffmann2004,OMill2008,Peng2010,Peng2014,Zheng2017}. For instance galaxies inhabiting denser environments usually present early type morphology and are brighter and redder than galaxies in low density regions. 
There is also a direct link between the main mechanisms that affect galaxies and their environment. Gas removal due to ram pressure in clusters, harassment leading to gas consumption by short and sequenced interactions, shock heating of infalling gas by the dark matter halos, are environmental quenching processes that shut down star formation \citep{GunnGott1972,Larson1980,Moore1996,DucBournaud2008,Peng2015}.
Tidal interactions and mergers between galaxies can also modify their morphology \citep{Toomre1972,HernandezToledo2005,Lofthouse2017} and trigger starbursts, therefore on average, interacting galaxies have enhanced star formation rates \citep{Kennicutt1998,Lambas2003,Alonso2004,Ellison2008,Patton2013,Pan2018}.
In this context galaxy groups which are compact, i.e. with close members and a low velocity dispersion, create a perfect scenario for galaxy-galaxy interactions and mergers and are excellent laboratories to study galaxy evolution. Nevertheless, even if groups are locally isolated there can be a contribution of the larger scale surroundings to the main properties of their galaxy members.

In the identification of galaxy groups there are a diversity of selection criteria that consider the local isolation of the system. For example, to select compact groups ``Hickson criteria'' \citep{Hickson1982}  states that there can not be significant neighbours within a concentric circle three times greater than the  smallest circle encompassing the geometric centres of the galaxies in the group. On the other hand, \citet{OMill2012} and \citet{Duplancic2015} selected isolated triple systems with no significant neighbours within a fixed aperture of $0.5\mpc$ from the centre of the system and with the same restriction on the radial velocity difference used to identify triplet members. For galaxy pairs the isolation is usually restricted to the absence of a third companion within the projected distance and radial velocity cut used to define the pair, and/or by requiring that galaxies in the pair sample do not belong to larger structures as groups or clusters of galaxies \citep[i.e.][]{Lambas2003,Ellison2010,Lambas2012,Patton2016}.
 \citet{Argudo2015} select catalogues  of  isolated galaxies,  isolated  pairs,  and  isolated  triplets by requiring no companions within  1$\mpc$ and radial velocity  differences $\Delta v <500 \kms$. 
 In \citet{Duplancic2018} we define homogeneous selection criteria of small and compact galaxy systems with two and up to six members in order to build catalogues suitable to compare main properties of pairs, triplets, and groups with four or more members. For these systems the isolation criteria accounts only for neighbours within a fixed aperture of $0.5 \mpc$ projected radius and a velocity difference of $\Delta V< 700 \kms$ with respect to the centre of the system.

 Regarding the study of  the large scale environment, for compact groups, a high percentage of the systems (50 to 70 percent) are found to be embedded in overdense regions such as clusters of galaxies or loose groups \citep[e.g.][]{RoodStruble1994,Barton1998,deCarvalho2005,AndernachCoziol2007,Mendel2011}. Nevertheless there are other works that found only about a 20 percent of these systems linked to larger structures \citep{Palumbo1995,Diaz-Gimenez2015}. 
Related to pairs and triplet of galaxies \citet{Argudo2015} found that most of these systems belong to the outer parts of filaments, walls and clusters. In this line  \citet{Tawfeek2019} study a sample of 315 isolated triplets and found that the dynamical evolution of these system is independent of their location in the Universe. For pair systems \citet{Tempel2015} find that the orientation of loose pairs correlates strongly with their host filaments showing a 25\% excess of aligned pairs compared to a random distribution. In this context \citet{Mesa2018} study the distribution of a sample of close galaxy pairs with respect to filament from the catalogue of \citet{Tempel2014b}, finding that 35\% of pair systems are closer than $1\mpc$ from the nearest filament axis. Moreover the authors found an excess of 15\% of pairs aligned with filaments, suggesting that the global environment may be influencing the accretion of galaxy pairs with a preferred orientation along the filament direction. 

It is important therefore to consider local as well as global environment in order to understand the physical processes that can drive galaxy evolution in small galaxy groups. There are different methods of measuring galaxy environment, the most popular are related to the distance to the n\textit{th} nearest neighbour \citep[e.g.][]{Balogh2004,Alonso2006,Ellison2010} and the number of significant neighbours within a fixed aperture \citep[e.g.][]{Kauffmann2004,Blanton2007,Berrier2011}. Other techniques to characterise the environment are the clustering of galaxy populations \citep{Peebles1973,Zehavi2002,Zehavi2005} and the identification of voids, sheets and filaments to describe the topology of the cosmic web and its relation with the properties of galaxies \citep{Ceccarelli2008,Lietzen2012}. 
 
In this context it is interesting to perform an environmental study considering a sample of small galaxy systems constructed under a homogeneous selection criteria. For this purpose in the present work we used the small galaxy system catalogue constructed in \citet[][hereafter D18]{Duplancic2018} where we identified compact and locally isolated systems with two or more galaxy members. For these systems we will study environment at larger scales than those considered for local isolation, through the implementation of diverse density estimators and also analyse the main properties of small system neighbour galaxies and their dependence with global environment.

This paper is organised as follows: in section \ref{data} we describe the galaxy catalogues used in this work. A study of the global environment of small galaxy systems is detailed is section \ref{analysis}. In section \ref{prop} we analyse the properties of neighbour galaxies of these systems at different scales. Finally in section \ref{conc} we present the main results of this work.

Throughout this paper we adopt a cosmological model characterised by the parameters $\Omega_m=0.3$, $\Omega_{\Lambda}=0.7$ and $H_0=70~h~{\rm km~s^{-1}~Mpc^{-1}}$.

\section{Data}
\label{data}

\subsection{Input Catalogue}
\label{parentsample}

The sample of galaxies were drawn from the Data Release 14 of Sloan Digital Sky Survey\footnote{https://www.sdss.org/dr14/} \citep[SDSS-DR14,][]{Abolfathi2018,Blanton2017}. This survey includes imaging in 5 broad bands ($ugriz$), reduced and calibrated using the final set of SDSS pipelines. The SDSS-DR14 provides spectroscopy of roughly two millions extragalactic objects including objects from 
the SDSS-I/II Legacy Survey \citep{Eisenstein2001,Strauss2002}, the Baryon Oscillation Spectroscopic Survey \citep[BOSS,][]{Dawson2013} and the extended-BOSS  \citep[eBOSS][]{Dawson2016}, from SDSS-III/IV.

In this work we consider Legacy survey and select extinction corrected model magnitudes which are more appropriated for extended objects by providing more robust galaxy colours. The magnitudes are k-corrected using the empirical k-corrections presented by \citet{OMill2011}. We restrict our analysis to galaxies with $r$-band magnitudes in the range $13.5<r<17.77$, the lower limit is chosen in order to avoid saturated stars in the sample and the upper limit corresponds to the limiting magnitude of the spectroscopic SDSS Main Galaxy Sample \citep[MGS,][]{Strauss2002}. We apply the apparent magnitude cut after Galactic extinction correction, in order to obtain an uniform extinction-corrected sample.
As we are aiming to perform an environmental study we want to reduce the survey edge effects, therefore we will consider only the contiguous area of the SDSS Legacy Survey. Also for spatial homogeneity between the different samples we will apply window and mask \texttt{polygon} files provided by NYU-VAGC\footnote{ics.nyu.edu/vagc/} \citep{Blanton2005} on our data.

\subsection{Small galaxy System Catalogue}
\label{catalogue}
In this work we use the catalogue of small galaxy systems presented in D18 constructed by using spectroscopic and photometric data from SDSS-DR14. For a detailed description of this catalogue we refer the reader to D18. 

Briefly the selection criteria is homogeneous in the identification of systems with a low number of members (two to six) populating  environments that promote galaxy-galaxy interactions and mergers. Galaxies in these systems are within the redshift range $0.05\le \rm z\le 0.15$ and have absolute r-band magnitudes brighter than $\rm M_{\rm r}$ $=-19$. Also, galaxy members within the system are close in projection in the plane on the sky ($\rm r_{\rm p}\le \rm 200 \kpc$) and have radial velocity differences $\Delta{\rm V}\le 500 \kms$. 
Systems are locally isolated by considering a restriction that prevent significant neighbours within a fixed circular aperture of 500~$\rm kpc$ radius with a radial velocity cut of 700 $\kms$. Also, their compacticity is similar to Hickson compact groups and galaxy members within the system have similar luminosities by considering the r-band absolute  magnitude difference between the brightest and faintest system members less than 2 magnitudes. Additionally galaxy systems in D18 have at least half of members with spectroscopic measurements being the fraction of spectroscopic galaxies equal to 90 per cent. 

The original catalogue of D18 reaches 10929 small galaxy systems in the entire Legacy survey area, nevertheless as in the present study we use geometrical constraints (see section \ref{parentsample}), the final sample used in this work comprises 9498 small galaxy systems from which 8705 are pairs, 715 triplets and 78 groups.

\subsection{Galaxy Tracers}
\label{Tracers}
As tracers of small galaxy system density environment we consider all spectroscopic galaxies in the Legacy Survey with redshift in the range $0.01<z<0.2$. These redshifts limits were selected to consider suitable  radial velocity cuts in the selection of small system neighbour galaxies. We remove from this sample galaxies in D18 catalogue in order to avoid contribution of system members in the environment density estimators.
These constraints yield to a sample of 546087 galaxies. On this parent sample we will apply different luminosity and radial velocity cuts in order to select significant neighbours that will be used in the computation of the diverse density estimators analysed in this work.  Also, for these galaxies we download the \texttt{galSpec} galaxy properties from MPA-JHU emission line analysis  based on the methods of \citet{Brinchmann2004}, \citet{Kauffmann2003} and \citet{Tremonti2004}, finding  99 percent of galaxies in the sample with spectroscopic measurements in MPA-JHU data. This information will be used to study the properties of neighbour galaxies of small systems.

\subsection{Random galaxies}
\label{random}
In the definition of environmental density estimators we will use samples of galaxies randomly distributed within the geometric area of SDSS. Therefore to reproduce the survey geometry  we use \texttt{mangle}\footnote{http://space.mit.edu/~molly/mangle/} \citep{Swanson2008} with the window and mask \texttt{polygon} files provided by NYU-VAGC\footnote{ics.nyu.edu/vagc/} \citep{Blanton2005}. Also we use Monte Carlo  algorithm  in order to select galaxies  with similar distributions of redshift and r$-$band luminosity  than those  of  the  tracer  samples.

\section{Small Galaxy Systems Environmental Analysis}
\label{analysis}

A direct way of studying the environment of systems is to calculate the density of surrounding significant neighbours. In every method used in this work we counted neighbours around the geometric centre of the system,  excluding galaxy members. Then we consider different environment estimators, as fixed aperture, distance to the n\textit{th} nearest neighbour, position with respect to cosmic filamentary structures and clustering properties through the correlation function.

\subsection{Fixed Aperture Environment Estimators}
\label{fixap}

\begin{figure}
  \centering
  \includegraphics[width=.45\textwidth]{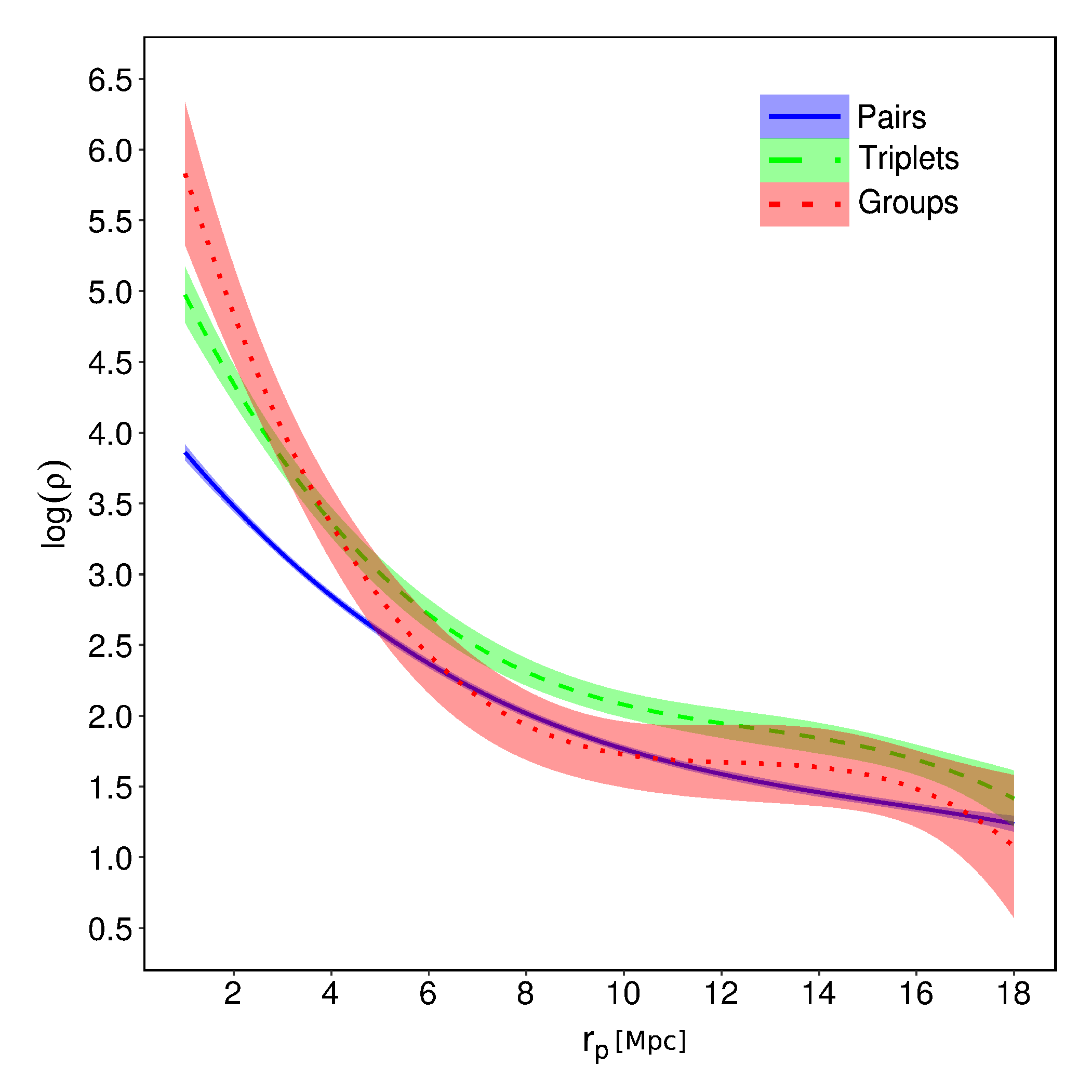}\hfill
  \includegraphics[width=0.45\textwidth]{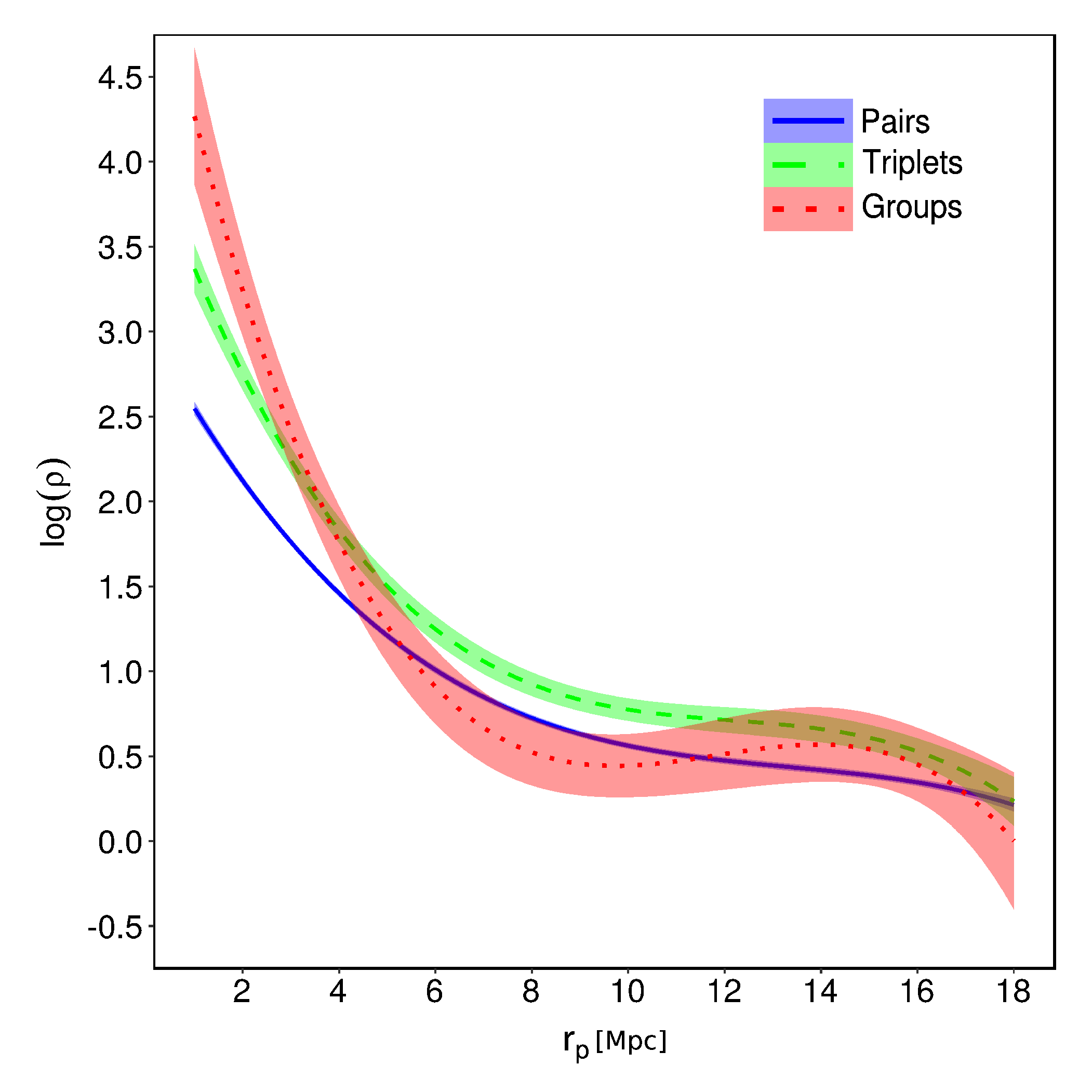}
\caption{ Density contrast as a function of distance to the system centre estimated by using fixed apertures (top) and annuli (bottom) counts for the sample of small galaxy systems, considering pairs (solid), triplets (dashed) and groups (dotted). We show a cubic spline fit to data and 95\% confidence intervals as shaded regions. The errors associated to pairs are very small given the sample size. } 
\label{aperture_annuli}
\end{figure}

To calculate fixed aperture density estimator, we consider as significant neighbours, galaxy tracers with absolute r-band magnitude $\rm M_{\rm r}\le -20.5$ and a radial velocity difference $\Delta \rm V<1000 \kms$ with respect to the small galaxy system centres. Also we use a sample of random galaxies with equal number of objects than significant neighbours. This sample was built as described in section \ref{random}.

In order to compute a cumulative density estimator we count galaxies within fixed apertures of $1 \mpc$ increasing radius. We test alternate bin widths finding that $1 \mpc$ bins provided the best tracer of small scale overdensities.  As an alternative to the former method we consider a differential estimator counting galaxies in annuli of 1$\mpc$ increasing inner and outer radius, rather than within fixed apertures. According to \citet{Muldrew2012} this technique removes the influence of local regions, better proving the influence of the larger scale environment. 

Then, we consider a scale up to $20 \mpc$ from system geometric centres and calculate the density contrast
$$\rho= {{N_{\rm gx} - \overline{N}_{\rm ran}} \over{\overline{N}_{\rm ran}}}$$
 where $N_{\rm gx}$ is the number of significant neighbours found in the aperture/annulus, and $\overline{N}_{\rm ran}$ was calculated by averaging the number counts of random galaxies within the aperture/annulus at the position of the system centres, i.e. is the mean number of neighbours that would be expected  if they were instead distributed randomly.

 In Fig. \ref{aperture_annuli} we present the results for overdensity estimators calculated by using number counts in increasing apertures and annuli, respectively. In these figures we show a cubic spline fit to the distribution of $\rho$ with respect to increasing radius, considering pairs, triplets and groups. The associated errors, represented as shaded regions, correspond to 95\% confidence intervals. For pair systems, errors are very small given the larger number of objects in the sample. 
 We notice that the curves diverge at distances smaller than 3$\mpc$, on these scales pairs have a general tendency of being surrounded by a lower number of significant neighbours than systems with a larger number of members. Triplets have intermediate overdensity values between pairs and groups, the latter are associated to denser environments.
The density contrast is similar at scales of about 5$\mpc$ irrespective to the number of members in small galaxy systems. From 5 to 10$\mpc$ the profile of triplets show a slight density increment with respect to pairs and groups, more evident from the annuli density profile. At scales larger than 10$\mpc$ pairs, triplets and groups present similar density contrast distributions.

 \subsection{Nearest Neighbour Environment Estimators}
\label{NN}
 The nearest neighbour approach studies the environment density by considering a variable scale estimator. Usually the surface density  parameter is calculated as 
$$\Sigma_{\rm n}={{\rm n}\over{\pi\ \rm r_{\rm n}^2}}$$ 

where n is the number of neighbours within a circumference with radius equal to $\rm r_{\rm n}$, the projected distance to the $\rm n^{\rm th}$ nearest significant neighbour. Defined in this way systems with closer significant neighbours are located in denser environments.

\begin{figure}
  \centering
  \includegraphics[width=.45\textwidth]{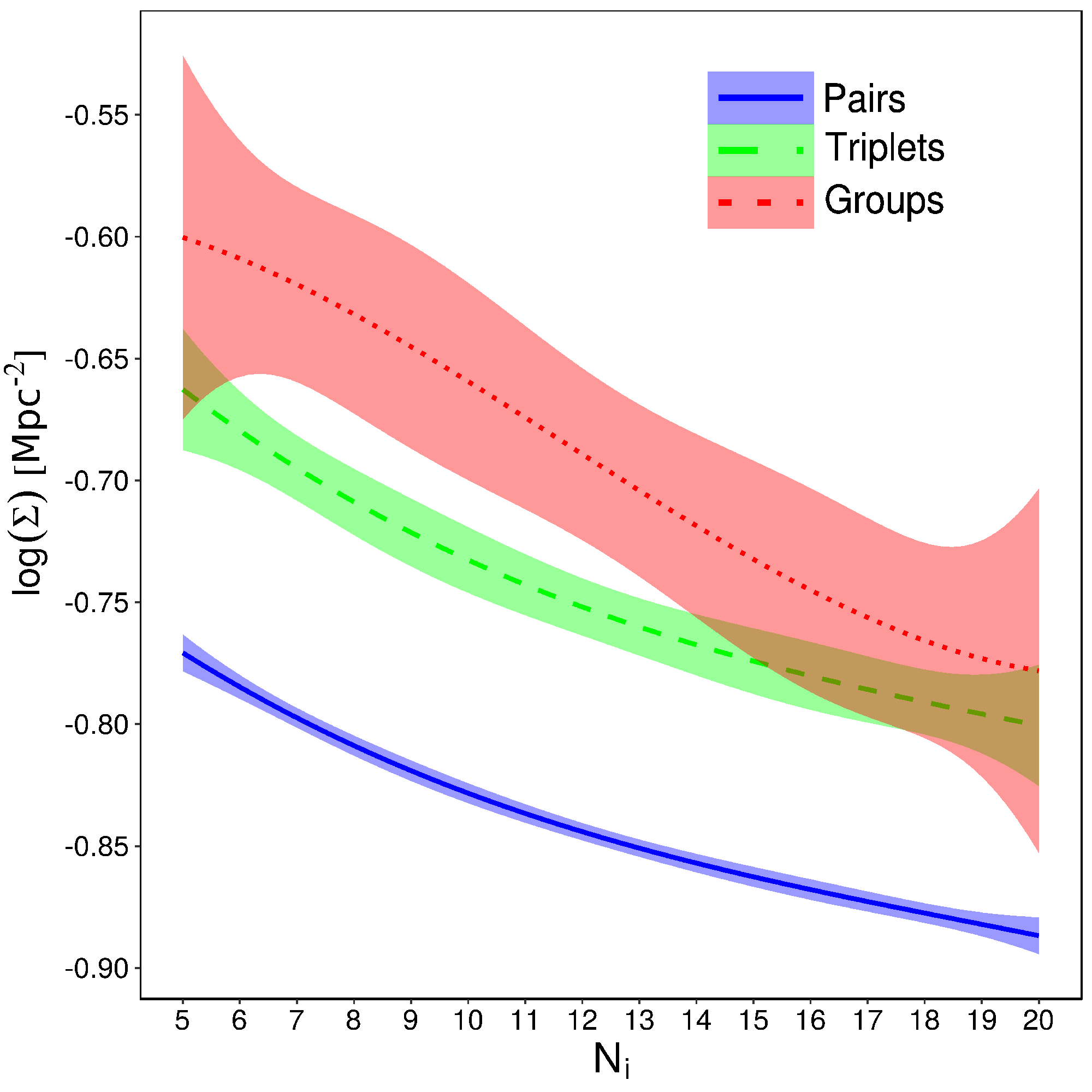}
\caption{$\Sigma$ parameter as a function of the $\rm n^{\rm th}$ nearest neighbour, of pairs (solid), triplets (dashed) and groups (dotted). We show a cubic spline fit to data and 95\% confidence intervals as shaded regions.} 
\label{sigma}
\end{figure}

In this work we calculate $\Sigma_{\rm n}$ with n=5,...,20 by considering as significant neighbour, galaxy tracers with  absolute r-band magnitude $\rm M_{\rm r}\le -20.5$ and with a radial velocity difference $\Delta \rm V< 1000 \kms$ with respect to the geometric centre of small galaxy systems. In Fig. \ref{sigma}  we show the values of $\Sigma$ for the different $\rm n^{\rm th}$ nearest neighbours, of pairs, triplets and groups.  In this plot $\Sigma$ limits correspond to a distance range of 2.5 to 3.5 $\mpc$ for the $5^{th}$ nearest neighbour and of 4 to 7 $\mpc$ for the $20^{th}$ nearest neighbour.

As in the fixed aperture density estimators this analysis indicates a difference between the environment density of pairs and groups with more members, but the nearest neighbour approach also highlights a difference when comparing triplets and groups. Triplets appear to be an intermediate systems between pairs and groups, preferring denser environments than pairs, but slightly less dense environments than groups. The density of triplets and groups is similar for neighbours farther than the $15^{\rm th}$ while galaxy pairs present lower densities for all the  nearest neighbour range.

\subsection{Distance to cosmic filaments}
To quantify the influence of large-scale structures on small galaxy systems, we consider the distance to their nearest filament ($d_{min}$). To this end we used the cosmic filament catalogue of \citet{Tempel2014b} where filamentary structures are represented through cylindrical segments of fixed 0.5$\mpc$ radius containing galaxy overdensities. In this catalogue filaments are traced by a set of points that define the filament axis, i. e. the spine.  For a detailed description of the catalogue see \citet{Tempel2014b}.

We consider the distance from the geometric centre of the system in order to avoid redshift space distortions arising  from  the  peculiar  velocities  of individual member galaxies. Also, we consider filaments longer than 10$\mpc$ as reliable filaments. Then we calculate $d_{min}$ as the distance from the geometric centre of each small galaxy system to the nearest filament spine point. 

\begin{figure}
  \centering
  \includegraphics[width=.47\textwidth]{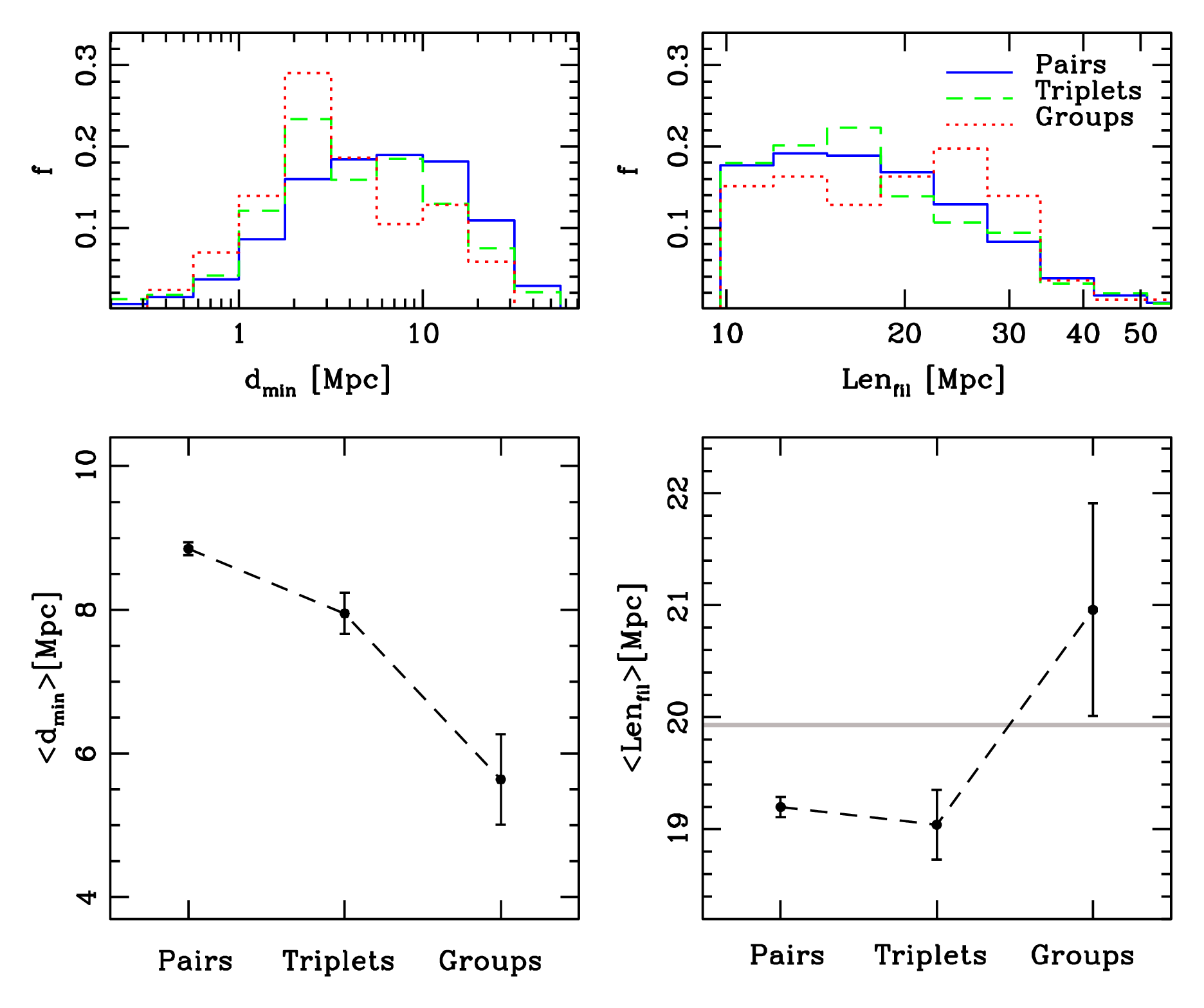}
\caption{Top: distributions of the distance to the nearest filament (left) and its length (right) for filaments associated with pairs (solid), triplets (dashed) and groups (dotted). Bottom: Mean value of the distance to the nearest filament (left) and filament length (right) for pairs, triplets and groups. The grey horizontal line represents the mean length of filaments in the sample under consideration. Error bars correspond to standard errors.} 
\label{dfil}
\end{figure}

In Fig. \ref{dfil} we show the distributions of the distance to the nearest filaments for pairs, triplets and groups. By definition small galaxy systems are locally isolated in a fixed aperture of 0.5$\mpc$ which is the radius of the filaments considered in this work, for this reason 98\% of systems in our sample are located at distances greater than 0.5$\mpc$ from nearest filament spine. In Fig. \ref{dfil} we also show the mean distance to nearest filament for pairs, triplets and groups, finding that pairs are located at average distance of 8.85$\pm$0.09, triplets at 7.95$\pm$0.29 and groups  5.64$\pm$0.63 from cosmic filaments. \citet{Kuutma2017}  consider that galaxies up to 10$\mpc$ from filaments are deep inside voids, therefore pairs and triplets in our sample are located in void environments while galaxy groups are more likely to reside in void walls. 
Several works \citep[e.g.][]{Barton2007,Alonso2004, Alonso2012, Ellison2010} have shown that galaxies in pair systems are preferentially located in group environments. Nevertheless, in this work, pairs, triplets and groups are locally isolated due to a strict isolation criteria considered in the identification of small galaxy system. This criteria excludes neighbours within a fixed aperture of 500$\kpc$ ($\Delta{\rm V}\le 700 \kms$). In \citet{Duplancic2015} we use these restrictions to isolate galaxy triplets finding that 95 per cent of the systems are at distances greater than $3\mpc$ from clusters, therefore the isolation is effective in the identification of systems far away from high density regions. In this context, pairs studied here are not only interacting galaxies but individual systems composed by two member galaxies and inhabiting low density environments.

In right panels of Fig. \ref{dfil} we explore the length of the nearest filament of small galaxy systems. We found that on average pairs are located closer to filaments of 19.02$\pm$0.09 $\mpc$ length, triplets to filaments with a length of 19.04$\pm$0.31 $\mpc$ and the length of filaments closer to groups are on average  20.96$\pm$0.95$\mpc$. Therefore groups are associated to larger filaments than pairs and triplets that reside closer to filaments of similar length. It is worth to notice that the mean length of filaments in the sample under consideration is 19.93$\pm$0.02$\mpc$ therefore, pairs and triplets are located close to filaments of shorter length than the mean while groups are closer to long filaments.
This result could be expected since longer filaments imply an enhancement of density therefore a higher fraction of elliptical galaxies can be found \citep{Mesa2018}. So, following the morphology-density relation \citep{Dressler1980} the more massive galaxy systems are expected to be preferentially located close to longer filaments.

\subsection{Clustering Measurements}
\label{CorrelationFunc}
In this section we aim to study the clustering around small galaxy systems. To this end we compute the cross correlation function between pairs, triplets and groups, each with respect to the sample of galaxy tracers described in \ref{Tracers}. The two-point correlation function is a statistical tool that is widely used in cosmological and large scale structure studies. It calculates the probability of finding two objects separated by a distance $r$, and quantifies how much it diverges from a random distribution. This yields information about the distribution of galaxies around the different systems under study, compared to the overall galaxy distribution on a wide range of scales. 

\begin{figure}
  \centering
  \includegraphics[width=.50\textwidth]{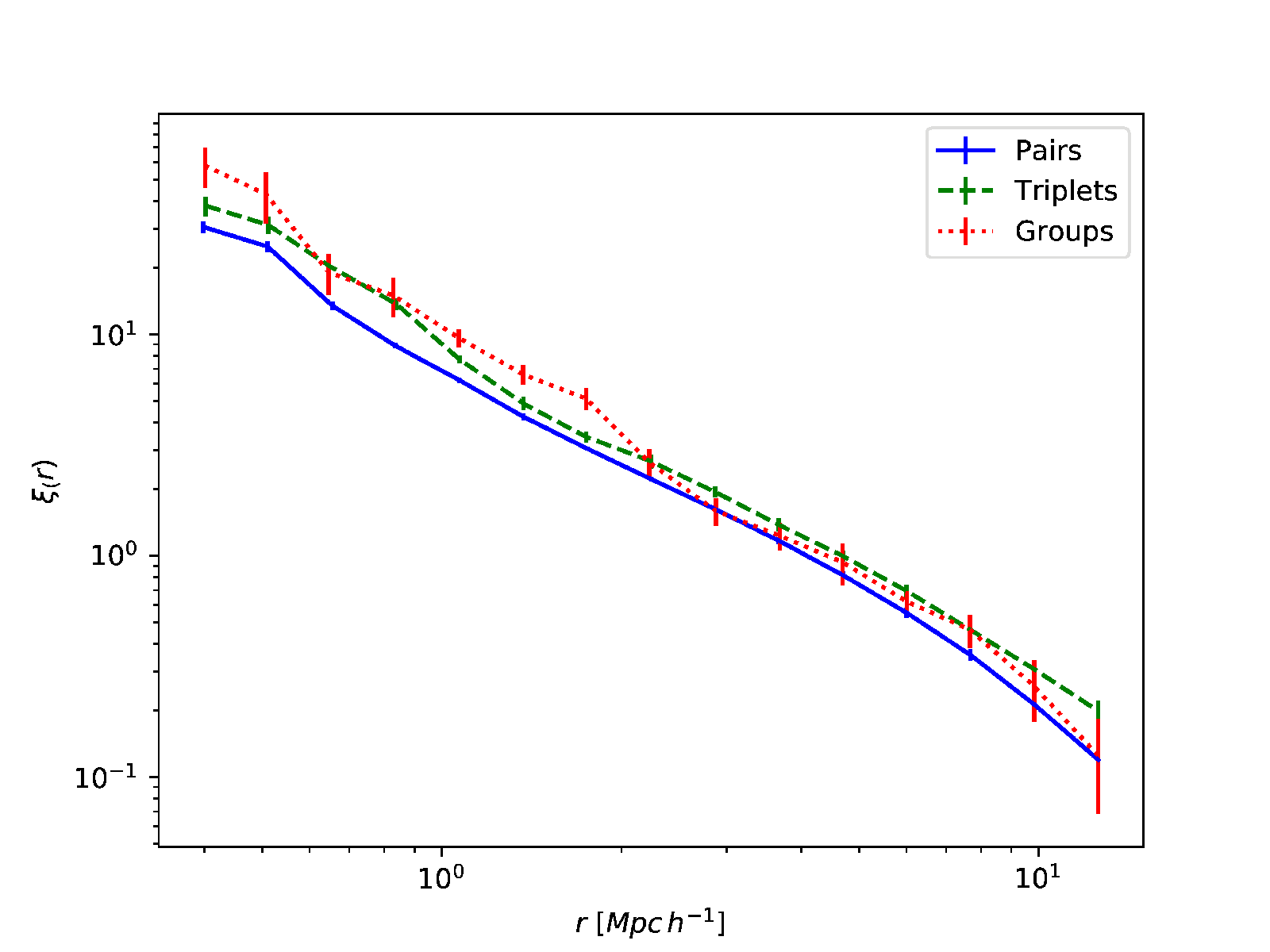}
\caption{Real-space cross-correlation functions for pairs (solid), triplets (dashed) and groups (dotted) in our sample. Errors were calculated with Jackknife resampling techniques. The errors associated to pairs are very small given the sample size. } 
\label{fc}
\end{figure}

We used the public \texttt{nbodykit}\footnote{https://github.com/bccp/nbodykit} \citep{Hand2018} to calculate  real-space Landy-Szalay estimator of the cross correlation function, adopting the cosmology introduced in section \ref{intro}, in the following form:

\begin{equation*}
\label{eqls}
\widehat{\xi}_{LS}(r)=1+\frac{D_{1}D_{2}(r)}{R_1R_2(r)}-\frac{D_1R_1(r)}{R_1R_2(r)}-\frac{D_2R_2(r)}{R_1R_2(r)},
\end{equation*}

where $D_iD_j$, $R_iR_j$, and $D_iR_j$, are the normalised number of data-data, random-random, and data-random pair counts respectively, for the different samples under consideration, binned as a function of the separation $r$. We used random samples as described in section \ref{random} considering 2, 20, and 200 times the size of the pair, triplets, and group samples, respectively. Errors were calculated with the Jackknife resampling technique by dividing the sample into 10 parts of equal number density. The use of 10 jackknife samples has been implement in different works to calculate associated errors of the correlation function \citep[e.g.][]{Ceccarelli2006,Zehavi2018}. Nevertheless we test 9 and 11 jackknife samples, finding no changes in our results.

The obtained cross correlation functions are shown in Fig. \ref{fc}. In agreement with the results discussed above, pairs inhabit environments of lower density when compared to triplets and groups, and on small scales ($< 3\mpc$) triplets appear to behave as an intermediate system.
Many studies have suggested that the galaxy correlation function has a change in its slope corresponding to the two-halo term according to the halo model and dark matter distribution \citep[e.g.][]{Neyman1952, Peacock2000, Seljak2000, Cooray2002, Jing2002, Sgro2013}. It is noteworthy that the three curves in Fig. \ref{fc} overlap in the two-halo term, i.e. on scales larger than $\sim $3 Mpc, but, on small scales there seems to be a difference in the amplitude of the cross-correlation of the systems under consideration. The differences in the slope and amplitude of the correlation functions on smaller scales suggest that pairs, triplets, and groups, are within halos of progressively greater mass. The fact that the curves overlap in larger scales is a consequence of the strong isolation criteria for the selection of the systems.

For the obtained correlation functions we performed a least-squares fit of a single power-law $\xi( r)= ( r/ r_0)^\gamma$, and obtained the fitting parameters and associated error shown Table \ref{table:1}. For the different samples there is a larger difference in the slopes than in the $r_{0}$ parameter. This indicates that groups are immersed in denser environments when compared to pairs and triplets.

The fitting parameters of the groups sample have larger errors than those of pairs and triplets. This is, on one hand, due to the smaller size of the sample, and on the other hand, because data appears to deviate from a simple power-law behaviour, i. e. the group correlation function shows a larger difference between the one- and two-halo terms. 
The parameters obtained for the groups are similar to the ones obtained by \citet{Sanchez2005} for the outer-region cross-correlation of 2dF groups and galaxies from the APM galaxy survey \citep[][]{Maddox1990} with a magnitude $b_{j}<20.5$; they obtained $r_{0}=4.82\pm0.3\ \mpc \rm \ h^{-1}$ and $\gamma=1.46\pm0.1$. Within error bars, these parameters are also consistent with our estimation for the triplets sample.

For galaxy pairs the obtained parameters are more similar to those of galaxy auto-correlation function of faint galaxies $\rm M_{\rm b_{j}} - 5\ \rm log\ h\simeq-18.4$; $r_{0}\simeq3.7\pm0.8\ \mpc \rm \ h^{-1}$ and $\gamma=1.76\pm0.1$ \citep[]{Norberg2002}. Furthermore, \citet{Zehavi2002} found that for blue galaxies $r_{0}\simeq4.02\pm0.25\ \mpc \rm \ h^{-1}$ and $\gamma=1.41\pm0.4$, the similarity of these values to the ones shown in Table \ref{table:1} of our pairs sample can be expected given the results from D18 showing that galaxies in these systems are statistically blue. Moreover the isolation criteria is also expected to be the reason for the slopes in Table \ref{table:1} being lower than those in the literature given a similar clustering length $r_0$.

\begin{table}
\centering
\begin{tabular}{||c c c c||} 
 \hline
  & $r_{0}$  & $\gamma$ \\ [0.5ex] 
  &$[\mpc \rm \ h^{-1}]$&\\
 \hline\hline
 Pairs & $3.85 \pm 0.12$ & $1.47\pm 0.03$ \\
 \hline
 Triplets & $4.44 \pm 0.25$ & $1.49\pm 0.05$ \\ 
 \hline
 Groups & $4.33 \pm 0.52$ & $1.64\pm 0.11$ \\ [1ex] 
 \hline
\end{tabular}
\caption{Power-law fitting parameters for galaxy-system cross-correlation function.  The fit was made with a least squares method.}
\label{table:1}
\end{table}

\section{Properties of neighbour galaxies }
\label{prop}

\begin{figure*}
  \centering
  \includegraphics[width=.90\textwidth]{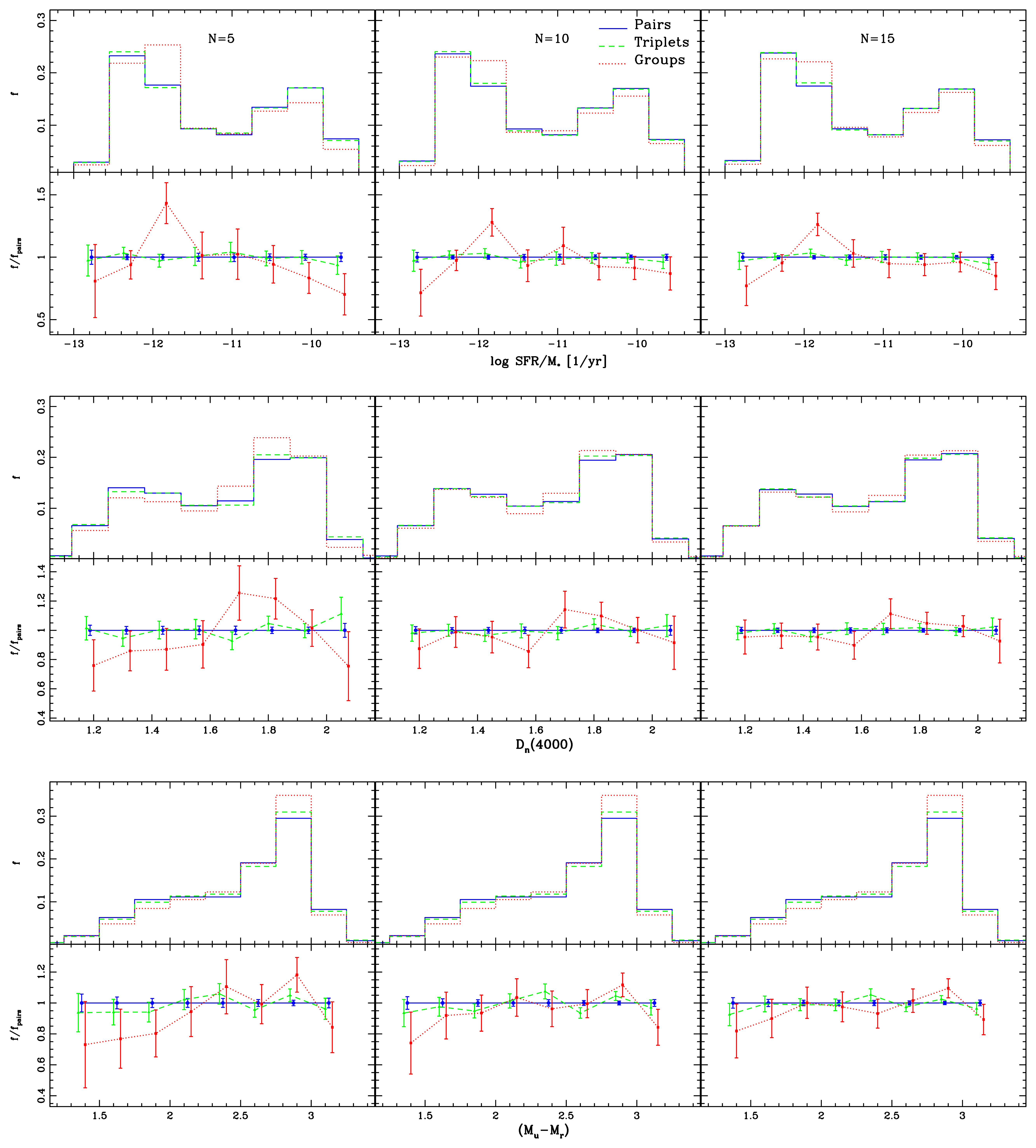}
\caption{ Normalised distributions of the properties of N nearest neighbour galaxies of pairs (solid), triplets (dashed) and groups (dotted) and quotient between the distributions of triplets and groups with respect to pairs. The error bars correspond to standard uncertainties. From top to bottom, specific star formation rate $\rm SFR/M_*$, $\rm D_n(4000)$ index and ($\rm M_u-\rm M_r$) colour.} 
\label{groupsprop}
\end{figure*}

In the following analysis we used galaxies brighter than $\rm M_{\rm r}=-20.5 $ from the tracers sample described in section \ref{Tracers} to study the properties of the N significant neighbour galaxies ($\Delta \rm V<1000 \kms$) of pairs, triplets and groups. We consider a variable scale to select significant neighbours based on the results of section \ref{analysis} which show that the nearest neighbour approach is more suitable to describe the global environment of small galaxy systems.

To study the main properties of galaxies, from the  MPA-JHU emission line catalogue we consider as a spectral indicator of the stellar population mean age the strength of the $4000$ \AA{} break ($\rm D_n(4000)$) defined as the ratio of the average flux densities in the narrow continuum bands 3850-3950 \AA{} and 4000-4100 \AA{} \citep{Balogh1999}. We also use the specific star formation rate parameter, $\rm log (\rm SFR/M_*$), as a good indicator of the star formation activity, according to \citet{Brinchmann2004}, and ($\rm M_{\rm u}-M_{\rm r}$) colour.  
In Fig. \ref{groupsprop} we show the normalised distribution of the properties of the 5, 10 and 15 nearest neighbours of pairs, triplets and groups, respectively. These values were selected in order to map a wide range of the nearest neighbour distribution, studied in section \ref{NN}.
It can be appreciated that nearest neighbours of group galaxies show an excess toward lower $\rm log(\rm SFR/M_*$) values with respect to neighbour galaxies of pair and triple systems. This difference is less evident in the $ \rm D_n(4000)$ index and ($\rm M_{\rm u}-M_{\rm r}$) colour distributions.
In order to compare and quantify the differences in the properties of neighbour galaxies of groups, triplets and pairs, we divide the fraction of the nearby galaxies of triplets and groups with the fraction of neighbours of the pair sample. The quotients are shown in the bottom panels of Fig. \ref{groupsprop}. 
It can be seen from this figure that groups tend to present lower fractions of star-forming, young stellar population, blue colour neighbour galaxies, compared to the neighbours of pairs. This tendency is more evident when considering the five nearest neighbours of the systems and dilutes to more distant galaxies. Moreover, objects near triplets and pairs present similar fractions.

In D18 we found that galaxies in small systems become less star-forming, with older stellar populations and with redder colours as the number of members in the system increases. The findings of this section indicates that the properties of the nearest neighbours reflect those of the member galaxies of small systems as found by D18, especially when considering the five nearest neighbours.

\section{Summary and Results}
\label{conc}

In this work we study the global environment of the sample of small galaxy systems constructed in D18. To this end we test several methodologies to characterise environment. We consider fixed aperture and nearest neighbours estimators. Also we calculate the position of the system in the cosmic web as well as the clustering of galaxies around the systems, through the cross-correlation function. We compare the environment of pairs, triplets and groups and the properties of neighbouring galaxies of these systems. The main results of these analysis can be summarised in the following items.

\begin{itemize}
    \item  When we consider a fixed aperture/annulus density estimator, a difference can be appreciated, at scales lower than 3$\mpc$, between the profile of pairs and the one corresponding to systems with a larger number of members. Pairs are associated to lower density environments than triplets and groups. The density contrast is similar at scales larger than 5$\mpc$ irrespective of the number of members in small galaxy systems. 

    \item The nearest neighbour approach also indicates a difference between the environmental density of pairs and systems with more members, but this density estimator also highlights a difference between triplets and groups. Triplets prefer denser environments than pairs, but slightly less dense environments than groups. Nevertheless, the density of triplets and groups is similar for neighbours farther than the $15^{\rm th}$ while galaxy pairs present lower densities for all the  nearest neighbour range.
    
    \item We consider the position of small galaxy systems within the cosmic web finding that pairs and triplets in our sample are located in void environments while galaxy groups are more likely to reside in void walls. Our findings also indicate that groups are associated to long filaments, while pairs and triplets are located close to filaments of lower length than the mean of filaments under consideration. Longer filaments have an enhancement of density and a higher fraction of elliptical galaxies, therefore the more massive galaxy systems are expected to be preferentially located close to longer filaments.
    
    \item Also, we explore the clustering around small galaxy systems and, in agreement with the results discussed above, the system-galaxy cross-correlation function shows that pairs inhabit environments of lesser density compared to triplets and groups, and on small scales ($< 3\mpc$) triplets appear to behave as an intermediate system.
    
    \item Additionally we study the properties of nearest neighbour galaxies of small galaxy systems. The neighbours of groups show a tendency to lower fractions of star forming, young stellar population and blue colour galaxies with respect to neighbours of pairs. This tendency is more evident when considering the five nearest neighbours of the systems and dilutes to more distant galaxies. Moreover, objects near triplets and pairs present similar properties.
\end{itemize}

The results of this work indicates that the properties of the nearest neighbours reflect those of galaxies in small systems as found by D18, i.e. a variation of the specific star formation rate, stellar populations ages and colours of galaxies, which become less star-forming, with older stellar populations and redder colours as the number of member galaxies in the system  increases.  This tendency is more important when considering the closest significant neighbours.

In D18 we suggest that repeated interactions may trigger transient phenomena as shocks and  activate  the star formation suppressing  mechanisms more efficiently in galaxy groups than in triplets and pairs. In this work we also find a difference in the global environment of small galaxy systems being groups related to denser regions than pair galaxies and triple systems. Therefore as a complement to the results found in D18 we suggest that the differences in the properties of member galaxies in small systems are not only related to the existence of an extra galaxy  but also to the large scale environment inhabited by the system.

\section{Acknowledgments}

We thank the referee for providing us with helpful comments that improved this paper.
This work was supported in part by the Consejo Nacional de Investigaciones Cient\'ificas y T\'ecnicas de la Rep\'ublica Argentina (CONICET) and Secretar\'ia de Ciencia y T\'ecnica de la Universidad Nacional de San Juan. Funding for the Sloan Digital Sky Survey IV has been provided by the Alfred P. Sloan Foundation, the U.S. Department of Energy Office of Science, and the Participating Institutions. SDSS-IV acknowledges support and resources from the Center for High-Performance Computing atthe University of Utah. The SDSS web site is www.sdss.org. SDSS-IV is managed by the Astrophysical Research Consortium for the Participating Institutions of the SDSS Collaboration including the Brazilian Participation Group, the Carnegie Institution for Science, Carnegie Mellon University, the Chilean Participation Group, the French Participation Group, Harvard-Smithsonian Center for Astrophysics, Instituto de Astrof\'isica de Canarias, The Johns Hopkins University, Kavli Institute for the Physics and Mathematics of the Universe (IPMU) / University of Tokyo, Lawrence Berkeley National Laboratory, Leibniz Institut f\"ur Astrophysik Potsdam (AIP),  Max-Planck-Institut f\"ur Astronomie (MPIA Heidelberg), Max-Planck-Institut f\"ur Astrophysik (MPA Garching), Max-Planck-Institut f\"ur Extraterrestrische Physik (MPE), National Astronomical Observatories of China, New Mexico State University, New York University, University of Notre Dame, Observat\'ario Nacional / MCTI, The Ohio State University, Pennsylvania State University, Shanghai Astronomical Observatory, United Kingdom Participation Group, Universidad Nacional Aut\'onoma de M\'exico, University of Arizona, University of Colorado Boulder, University of Oxford, University of Portsmouth, University of Utah, University of Virginia, University of Washington, University of Wisconsin, Vanderbilt University, and Yale University.
\bibliographystyle{mnras.bst}
\bibliography{main}{}

\label{lastpage}

\end{document}